\documentclass[pra,showpacs,uplatex]{revtex4}

\usepackage{amsmath}
\usepackage{bm}
\usepackage[dvipdfmx]{graphicx}
\usepackage[dvipdfmx]{color}
\usepackage[dvipdfmx]{epsfig}
\usepackage{float}
\usepackage{ulem}
\usepackage{color}

\begin{document}

\title{Domain wall dynamics in the spinor Bose-Einstein condensates}

\author{Hiroshi Kuratsuji}
\email{kra@se.ritsumei.ac.jp}
\address{Office of professor emeritus, Ritsumeikan University-BKC, Kusatsu City, Shiga, 525-8577, Japan}

\date{\today}

\begin{abstract}
A dynamical theory is presented for the domain wall in the spinor Bose-Einstein condensate of "effective"one-dimension.   
The formulation is based on the time-dependent Landau-Ginzburg (LG) theory written in terms of 
the spinor order parameter.  The procedure is carried out in such a way that an adiabatic change of the collective coordinates; 
the kink position as well  as the phase function built in the spinor, is adapted to  the canonical 
 term (alias geometric phase) in the LG Lagrangian.  Two problems are discussed:  
The first one is to explore  the translational  motion of the kink center in  the presence of the pinning potential, which results in the quantum potential problem. 
 The other one concerns the {\it steering} of the domain wall 
by modulating the externally driven magnetic field assisted with dissipation. 
The present attempt would  shed light on the spinor Bose- Einstein condensates from a novel viewpoint. 
\end{abstract}

\maketitle

\section{Introduction}

One of the major topics in condensed matter physics is to study  the defects in quantum condensates.Among others, 
the vortex and the domain wall are typical objects \cite{LL1,LL2}.  
The domain wall (abbreviated as DW),   which was pioneered by Bloch \cite{Bloch},  
is  known to occur in wide class of materials showing up ferromagnetic structure\cite{Malozemoff}.   
The topics is still  in the limelight from various points of view (see e.g. \cite{Murugesh,Zhu,Natt,Goussev,Guru}).   
Furthermore to be mentioned is that it covers even beyond original  magnetic  material, 
 for example, the occurrence of DW in the superfluid He3A  \cite{He3A} 
 and the nematic liquid crystal \cite{LC}. 
 Apart from the ferromagnetic and similar materials, the DW is also expected to occur in the Bose-Einstein condensate (BEC) 
 exhibiting  the spin structure.  Indeed the DW has been recently investigated  for the  BEC of two components \cite{Filatrella}. 
 The domain wall was also studied in the spinor BEC in optical lattice model\cite{Li1} which was investigated 
  in the general context of the magnetic solitons \cite{Pu,Li2}.

 In this letter we put forward a theory for the dynamics of DW  in the spinor BEC, which 
 is characterized by the spin texture\cite{Ho1,Ho2}.  Specifically, by reducing the problem to  the one-dimensional 
 system, we discuss the following two topics: The first topics is to address the dynamics of the DW without external driven force, namely, 
 spontaneous translational motion of the DW leading to the problem as if a particle dynamics. 
The other  is to explore the motion of the center of DW in the presence of the time-modulated external field as well as 
the dissipative effect; which  is described under the concept of {\it steering} (or control).  

The starting point is the  Landau-Ginzburg (LG) type Lagrangian which  is designed in terms of the spinor oder parameter 
\cite{Harada,HK1}. The crux lies  in extracting  the motion of the collective coordinates that are incorporated in 
the kink profile, namely, the center of the kink as well as the phase function built in the spinor order parameter. 
The procedure is carried out such that  the adiabatic change of these  collective degrees  is adapted to the canonical term \cite{HK2}  
in the LG Lagrangian, 
which is a variant of the geometric phase. 
 Here we note that the essentially  different ways are  required to manipulate the canonical term for the wo topics,  as is seen 
  from the discussion below.

\section{Preliminary}

We begin with introducing the order parameter for the spinor BEC. Although it is possible to 
consider the condensate for the general spin, we shall here take up the simplest case, 
namely, a two-component  wave parametrized by angular variables $ (\theta, \phi) $  (see e.g. \cite{LL3}) 
\begin{equation}
\Psi = F \left(
\begin{array}{cc}
\cos\frac{\theta}{2}  \\
\sin\frac{\theta}{2} \exp[-i\phi] 
\end{array}
\right) 
\equiv F \psi, 
\label{spinor}
\end{equation}
where $ F $ represents the magnitude of the condensates.  

The Lagrangian (density) that  governs the order parameter introduced above is 
given as  \cite{Feynman,HK3} 
\begin{equation}
L = \frac{i\hbar}{2} (\Psi^{\dagger} \dot \Psi - {\rm c.c.} ) - H(\Psi^{\dagger}, \Psi) 
( \equiv  L_C - H ) 
\end{equation}
(dot means $ \frac{\partial}{\partial t} $).  $ L_C $ is called the canonical term hereafter.  
As for the Hamiltonian, it is most natural to choose the simplest form, which  may be 
given by  the sum of the kinetic energy and potential energy: 
\begin{equation}
H = \frac{\hbar^2}{2m}\nabla\Psi^{\dagger}\nabla\Psi + V(\Psi^{\dagger}, \Psi)
\end{equation}
with $ m $ being the mass of the constituent atom composing the spinor BEC. 
In what follows we consider the case that the magnitude of the condensate is fixed to be $ F = F_0 $. 
This means that the condensation is describe by the Landau type potential: $ V(F) \sim (F^2 - F_0^2)^2   $. 
Hence  the angular field $ (\theta, \phi) $  is allowed to be the dynamical degree of freedom. 

In terms of the angular variable, the Lagrangian is written in the form
\begin{align}
L_C & =  \frac{\hbar}{2}F_0^2 (1-\cos\theta)\dot\phi,    \nonumber \\
H & =  \frac{\hbar^2}{2m}F_0^2\{\nabla \theta)^2 + \sin^2\frac{\theta}{2}(\nabla\phi)^2 \} +V . 
\end{align}
The equation of motion for the order parameter for the spinor BEC is derived by applying the 
variational principle; $ \delta \int L dxdt = 0 $: 
$$
{\partial \theta \over \partial t} =  {1 \over \sin\theta} {\delta H  \over \delta \phi}, 
~~{\partial \phi \over \partial t}  = 
 -{1 \over \sin\theta}{\delta H \over \delta \theta} . 
 $$
 Here for later use, the spin variable is introduced: 
 \begin{align}
S_x& (= \Psi^{\dagger}\sigma_x \Psi) = \sin\theta\cos\phi , ~~S_y (= \Psi^{\dagger}\sigma_y \Psi )= \sin\theta\sin\phi, 
\nonumber \\
S_z &  (= \Psi^{\dagger}\sigma_z \Psi)  = \cos\theta . 
\end{align}

\section{The construction of the kink solution}

In what follows the argument is restricted to the one-dimensional system in  which the spin vector depends only on the coordinate along  $ x $ -direction. 
The procedure is the same as the model that was discussed in the previous paper \cite{Harada}, namely, the transverse coordinate $ (y, z ) $ 
is integrated over to reduce to the one dimensional system.   This procedure  is similar to the model that was studied for the current through a narrow channel \cite{Langer}. 

As for the potential term $ V $, it is assumed to consist of 
the self interaction among the order parameter and  the other terms coming from the pinning 
and externally controlled field,  namely 
\begin{equation}
V = g F_0^4 (\Psi^{\dagger}\sigma_z \Psi)^2  + V_{extra} 
\end{equation}
and the Hamiltonian can be written in the form $ H = H_0 + V_{extra} $: 
\begin{align}
H_0  & =  \frac{\hbar^2}{2m}F_0^2\{(\frac{d\theta}{dx})^2 + \sin^2\frac{\theta}{2} (\frac{d\phi}{dx})^2\} + gF_0^4 \cos^2\theta . 
\label{H0}
\end{align}
The last term comes from the interaction between the spin of the condensate which is a modified form 
used in the paper \cite{Ho1}, for which the coupling constant $ g $ is chosen to be positive. 
The second term in the kinetic energy, which is expressed as the gradient of the phase $ \phi $ means the fluid 
kinetic energy, which plays a key role in determining the kinetic energy of the DW 
as will be shown below. 
Now as the first step we construct the specific solution in the static limit; namely, we look for 
a solution that satisfies the variational equation for $  H_0 $, leading to the 
coupled equation 
\begin{eqnarray}
\frac{d^2\theta}{dx^2} - \sin\theta(\frac{d\phi}{dx})^2 - \frac{mgF_0^2}{\hbar^2}\sin^2\theta  & =  & 0, \nonumber \\
\frac{d}{dx}\big(\sin^2\frac{\theta}{2} \frac{d\phi}{dx}\big)  & =  & 0 . 
\end{eqnarray}
From the second equation one has the "constant of motion" $  \sin^2\frac{\theta}{2} \frac{d\phi}{dx}
= \rm{constant} $, which can be  set to be zero. Hence we have an integral of motion: 
\begin{equation}
 \big(\frac{d\theta}{dx})^2 = \frac{1}{\lambda^2} \sin^2\theta 
\end{equation}
leading to 
\begin{equation}
\cos\theta (\equiv S_z) = \tanh[\frac{x}{\lambda}], ~~
\sin\theta = \frac{1}{\cosh\frac{x}{\lambda}},  
~~\phi = {\rm constant} 
\label{kink}
\end{equation}
with $ 1/\lambda^2 = \frac{2mgF_0^2}{\hbar^2} $ (the inverse of the width of the DW). 
 Thus one sees that the spin behaves such that 
$ S_z(x)  $ changes from $ \ - 1 $ for $ x \rightarrow -\infty $  to  $ + 1 $ for $ x \rightarrow  + \infty $, which shows 
the width by $ \lambda $ near $ x = 0 $ that corresponds to the coherence length. This way, 
Eq.(\ref{kink}) just represents the kink solution (Fig1).  This is the starting point for the dynamical problem that will follow. 
\begin{figure}[htb]
  \hspace{10mm}
  \includegraphics[width=45mm]{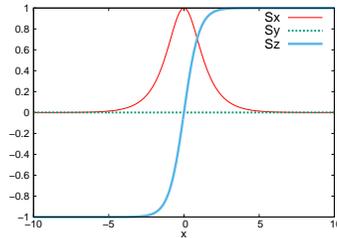}
  \vspace{11mm}
  \begin{center}
    \caption{(Color Online).
Profile of the kink. The case $ \phi = 0 $ is adopted.  }
    \label{fig:Lam1.0}
  \end{center}
\end{figure}

\section{Motion of the domain wall}

Now we address the first topic; the behavior of the domain 
wall under the interaction with a pinning immersed in  the condensate. 
  Concretely speaking, we are concerned with the translational motion for the kink center. 

To carry out this, we adopt the "adiabatic procedure" to extract the kinetic energy for the kink center. That is, 
the motion of the kink is incorporated  by shifting $ x \rightarrow x - R(t) $ with $ R(t) $ denoting the coordinate 
of the 
center of the kink,  hence  the kink profile is expressed as the form $ \cos\theta(x- R(t)) $. 
Namely, the profile of the DW keeps the same form as in the static case, which serves as a basis 
for the adiabatic approximation.  
One more point  is the choice of the phase function $ \phi $:  
The immediate form is  
 the form $ \phi(x  - R(t)) $ that is similar to  the DW profile $ \theta(x) $, which 
 we adopt  in what follows. Here note  that  the evolution of $ \phi(x-R(t)) $ has no 
 connection with that of $ \theta(x- R(t)) $, which is a simple consequence that the field variable $ \phi(x) $
 and $ \theta(x) $ are independent  of each other.
 This  parametrization is the same as the one used in the vortex
\cite{HK2,HK3}.  However, as is seen below, the procedure is quiet different from the vortex case, because 
the case of vortex deals with two-dimensional defect.

 Noting the parametrized form of the profile of the kink, it follows that 
\begin{equation}
\frac{\partial \phi}{\partial t} = -\dot R\frac{\partial \phi}{\partial x} , 
\end{equation}
hence the canonical term is written as 
\begin{equation}
L_C = - \frac{\hbar F_0^2}{2}\int (1-\cos\theta)\frac{d\phi}{dx} \dot R \equiv  P \dot R . 
\end{equation}
From this expression, $ P $ appears to be the momentum conjugate to $ R $, as will 
be verified below. 

We now have the  problem  of how to control  the derivative $ \frac{d\phi}{dx} $; 
to  carry out this, we write  the fluid kinetic energy, which is denoted as  $ K_{\phi} $. 
Taking account of the feature that both $ P $ and $ K_{\phi} $ are written in terms of the derivative  $ \frac{d\phi}{dx} $,  we set 
the following  Ansatz for this derivative: 
\begin{equation}
\frac{d\phi}{dx} = C\rho(x) . 
\end{equation}
The quantity $ C $ represents the unknown parameter together with the {\it form factor} 
$ \rho $ which may be chosen appropriately.  Thus we have 
\begin{equation}
P = -C\frac{\hbar F_0^2}{2}\int^{\infty}_{-\infty}  (1- \cos\theta)\rho dx = -\frac{\hbar}{2}F_0^2CI . 
\label{P} 
\end{equation}
On the other hand, as for $ K_{\phi}  $,by noting the relation (\ref{P}) it turns out to be 
\begin{align}
K_{\phi} & = \frac{\hbar^2C^2}{2m} F_0^2\int \sin^2\frac{\theta}{2}(\frac{d\phi}{dx})^2 dx  \nonumber \\
  & 
  = \frac{\hbar^2}{2m} C^2F_0^2J, 
\end{align}
which is expressed in the form of usual kinetic energy: 
\begin{equation}
K_{\phi}  = \frac{1}{2}\big[\frac{J}{4mF_0^2I^2}\big] P^2  \equiv \frac{P^2}{2M_{eff}} . 
\end{equation}
The two integral $ I,J $ are  calculated by choosing the form factor $ \rho(x) $: 
For example we choose $ \rho(x) = \{\cosh(x/\lambda)\}^{-1}  $, for which the effective mass is 
estimated as  $ M_{eff}  \propto m \times F_0^2\lambda $, which  is the order of the atomic mass $ m $. 

Hence the effective Lagrangian for the DW is given as 
\begin{equation}
L_{eff} = P\dot R - H_{eff} \equiv P\dot R - \big(\frac{P^2}{2M_{eff}} + V_{extra}\big), 
\end{equation}
which leads to the canonical equation of motion: 
\begin{equation}
\frac{dR}{dt} =\frac{P}{M_{eff}}, ~~ \frac{dP}{dt} = - \frac{\partial V_{extra}}{\partial R}. 
\end{equation}
In this way, we have arrived at the one-dimensional potential problem. 

\subsection{Pinning potential and its implication}
Next we evaluate  the pinning potential, for which it is plausible to adopt a magnetic 
origin; namely,  let us suppose a model designed by the local magnetic field located  at several positions 
$ x= l_i $; 
for example, which may be fabricated  by the inverse-Faraday effect; the induced 
magnetic field caused by the circular polarized light \cite{Pershan}.  Then the pinning potential can be written as  the 
coupling with the magnetic spin possessed by the spinor BEC in the form: 
 $  V_{pin} = \sum_i  \mu_i \hat\sigma_z \delta(x- l_i) $, where $ \mu_i $ is 
the magnetic moment multiplied by  a  local magnetic field and $ \hat \sigma $ denotes 
the Pauli spin. 

We suppose that  there are two magnetic origins at $ x=0, l $; that is  
$V_{pin} = \mu \{\hat\sigma_z \delta(x) - \hat \sigma_z \delta(x-l)\} $
which leads  to the extra potential 
\begin{align}
V_{extra} & = \Psi^{\dagger} V_{{\rm pin}} \Psi \nonumber \\
  & = \mu\big[ \tanh\lambda (R) - \tanh\lambda  (R- l)\big] . 
\end{align}
 To be noted here is that the potential incorporates 
the two parameters $ (\mu, l ) $.  Specifically we are concerned with the parameter $ \mu $: 
This may regarded as the height of the potential (in what follows we consider the case $ \mu > 0 $) 
which may be controlled externally.  To examine the effect caused by the potential, we shall modify 
the potential profile $ \tanh $ by replacing  a rectangular barrier.  By this modification 
the essential physical content may not be lost and the following  procedure 
may be valid if the parameter $ \lambda $ is regarded as small compared with $ l $.

Turning to quantum mechanics, we have a problem of evaluating the  transmission of the 
"particle" (the kink center)   to the potential barrier: We borrow the well known result: 
Namely, let us suppose the "particle" coming from the left with the energy $ E $ which 
satisfies $ E> \mu $ (that means above the barrier), then  the transmission rate 
becomes \cite{LL3}
\begin{equation}
T = \frac{4E(E-\mu)}{E(E-\mu) + \mu^2 \sin^2\big[ \sqrt{2M_{eff}(E-\mu)}\frac{l}{\hbar}\big] } . 
\nonumber 
\end{equation}
From this expression we see that the transmission becomes maximum at  $
\sin \big[\sqrt{2M_{eff}(E-\mu)}\frac{l}{\hbar}\big] = 0 $, which is called the 
{\it resonance } transmission. For the fixed value for the incident energy $ E $, 
the resonance position may be tuned  by controlling the height parameter $ \mu $ 
which may be manipulated properly. 

{\it  A possible observable effect }: 
Apart from the quantum mechanical phenomenon mentioned above it is possible to consider 
the quasi-classical effect that is caused by a peculiar nature of the rectangular shape of 
the potential: Specifically we take up  a single pinning leading to a potential step,  
which gives rise to an impulsive force to the pinning object. This effect may be utilized for a potential application:
 for example, a switching device with use of the macro fabrication.

\section{Steering of the DW under the modulated magnetic field}

We now turn to the second topic:the steering of the domain wall.  
The essential point is that we are interested in  the effect of the time-dependent external 
field.  Taking into account this feature,  we  need to adopt a completely different procedure for the canonical term  
from the one developed in the  previous section.  Keeping mind in  translational motion of the DW, we propose the following form  of the 
phase function $ \phi(x,t) $: 
\begin{equation}
\phi(x,t) = \eta(x-R(t)) \Phi(t) . 
\label{eta}
\end{equation}
Here in contrast to the previous case, $ \Phi(t) $ is a time-dependent dynamical variable, 
whereas $ \eta $, which incorporates the translational degree, is given as an 
Ansatz from outset in such a form that it is reflection symmetric function and with  boundary condition 
$ \eta(\infty) = \eta(-\infty) = 0  $
that has the  peak at the origin.  Furthermore $ \eta $ is a slowly varying function. A typical function is 
supposed to be  $ \eta(x) \propto \exp[-\varepsilon x^2] $.  We note that the form (\ref{eta}) may be regarded 
a modified form used in the other class of problem, 
such as soliton quantization\cite{Tjon}.

Now let us reduce the canonical term in the action function, which consists of two terms: 
$ S_C = S_C^1 + S_C^2 $: 
\begin{align}
S_C^1 & = \frac{\hbar F_0^2}{2} \big[ - \int^{\infty}_{-\infty}  (1-\cos\theta) \frac{d\eta}{dx} \dot R \Phi dxdt \nonumber \\
 S_C^2 & =  \int^{\infty}_{-\infty}  (1-\cos\theta)\eta \dot\Phi dxdt \big] . 
\label{sum}
\end{align}
By integrating $ S_C^1 $  by part with respect to $ x $, it follows that 
$$
 -\big[(1-\cos\theta)\eta \big]^{\infty}_{-\infty}  + \int \frac{d}{dx}\{1 - \tanh(\frac{x-R}{\lambda})\} \eta(x-R)dx,  
$$
where the first term vanishes owing to the boundary condition for $ \eta $, and  noting  that the second term 
$ \frac{d\tanh x/\lambda }{dx} = \frac{1}{\cosh^2 \frac{x}{\lambda} } $ is sharply peaked at the origin, 
 then $ S_C^1 $ can be approximated as 
\begin{equation}
S^1_C \simeq   -\frac{\hbar F_0^2}{2} \int  \eta(0) \dot R \Phi dt . 
\label{LC} 
\end{equation}
As for the second term $ S_C^2 $, it  turns out by using the partial 
integration for the time integration: 
\begin{align}
S_C^2 & \propto  \int \big[ \frac{\partial}{\partial x}\{(1-\cos\theta)dx\} \big] \dot R \Phi dt \nonumber \\
    & = \big[(1-\cos\theta)\eta\big]^{\infty}_{-\infty} \times \dot R\Phi dt, 
\end{align}
which vanishes owing to the boundary condition for $ \eta $.  Thus we finally obtain for the 
effective Lagrangian for the canonical term: 
\begin{equation}
L_C^{eff} = \frac{\hbar F_0^2}{2} \eta(0) R \dot \Phi, 
\label{canon}
\end{equation}
where use is made of $ - \dot R\Phi \equiv R\dot\Phi $ up to the total derivative. 
As the next problem, we need to evaluate the fluid kinetic energy  $ K_{\phi} $ in the Hamiltonian $ H_0 $. However 
the argument about this will be put  aside and will be discussed briefly later. \\

Now we treat the effective Hamiltonian.We suppose the coupling of the spin component inherent in the BEC 
 with an external magnetic field, namely, 
\begin{align}
H_{eff} =   \int \Psi^{\dagger} {\bf \sigma}\cdot {\bf H}(t) \Psi dx.   
\label{Heff}
\end{align}
Here the magnetic field is arranged to be the rotating field in $ (x,y) $ plane: 
$ {\bf H}(t) = (b(t)\sin\alpha(t), b(t) \cos\alpha(t), 0) $ where the amplitude $ b $ is allowed to be 
modulated in time.  The integral of  (\ref{Heff}) is written as 
\begin{equation}
H_{ext} =  bF_0^2 \int \sin\theta \cos(\eta \Phi(t) - \alpha(t))dx. 
\end{equation}
The integral may be carried out approximately; noting that $ \sin\theta = {\rm sech}(x/\lambda) $ 
is a function peaked at $ x = 0 $, it may be regarded as if the delta function and hence the 
the above integral is approximated to give  
$ H_{ext} = bF_0^2 \cos(\eta(0)\Phi - \alpha(t)) $. 
In this way, the effective Lagrangian in the presence of the externally driven magnetic 
field 
\begin{equation}
L_{eff} = AR \dot \Phi - k \cos(\Phi - \alpha(t)), 
\label{Leff}
\end{equation}
 where we put $  A \equiv \frac{\hbar F_0^2}{2}, k \equiv bF_0^2 $ and use the scaling $ \eta(0)\Phi \rightarrow \Phi $. Note that  
 $ R $ and $ \Phi $ looks like canonical variable each other, but this feature will be 
 discussed briefly at the end of this section.

 {\it Dissipative effect} :  
In order to manipulate  the coupled dynamics for $ (R,\Phi) $ using the geometric term $ L_C $ 
properly,  it may be efficient to take into account  the dissipative effect which is described by 
the dissipation function (e.g. \cite{LL4,Yabu}), 
\begin{align}
F&  =  \int f \vert \frac{\partial \psi}{\partial t} \vert^2dx = fF_0^2 \int \{ (\frac{\partial \theta}{\partial t})^2 
+ \sin^2\frac{\theta}{2} ( \frac{\partial \phi}{\partial t})^2\} dx . 
\label{dissipation} 
\end{align}
In the second term of (\ref{dissipation}), only the term proportional to $ \dot\Phi^2 $  is kept, because 
the other terms are of higher oder in $ (R,\Phi) $ together with its derivatives  However these are irrelevant to the 
present procedure. Then  we write 
\begin{equation}
F = \frac{1}{2}(B\dot R^2 + D\dot\Phi^2), 
\end{equation}
where $ B, D $ are calculated as 
\begin{align} 
\frac{1}{2}B & \equiv  fF_0^2 \int (\frac{d\theta}{dx})^2 dx = fF_0^2 \frac{2}{\lambda},  \nonumber \\
~~\frac{1}{2}  D &  \equiv  fF_0^2  \int \sin^2\frac{\theta}{2} \eta^2 dx  = fF_0^2 \int_{-\infty}^{\infty} \eta^2(x)dx . 
\end{align}
Using this dissipation function the coupled equation of motions  for $ R $ and $ \Phi $ is obtained as \cite{LL4}
\begin{equation}
\frac{d}{dt}\big(\frac{\partial L_{eff}}{\partial \dot X}\big) - \frac{\partial L_{eff}}{\partial X} 
= \frac{\partial F}{\partial \dot X}
\end{equation}
for $ X = (R, \Phi) $.  We will  analyze this coupled equation for two cases separately. 

{\it Case (I)}: The case that the amplitude $ b $ (and hence $ k$)   is kept constant, for which 
we have 
\begin{equation}
A\dot R -  k \sin(\Phi - \alpha(t)) 
 =   D\dot \Phi,  ~  -A\dot \Phi   =  B\dot  R. 
\end{equation}
By eliminating $ R $, we have 
\begin{equation}
(D+\frac{A^2}{B})\dot\Phi = -k\sin(\Phi - \alpha(t)), 
\end{equation}
which  can be analyzed as follows: by introducing $ \tilde\Phi = \Phi - \alpha $. In the adiabatic scheme, we can 
put $ \tilde\Phi \simeq 0 $, hence we approximate 
\begin{equation}
\frac{d}{dt}(\tilde\Phi) + \frac{k}{d}\tilde\Phi = -\dot\alpha
\end{equation}
by  putting $ d = D+\frac{A^2}{B} $.  We find a special solution for $ \Phi $, 
\begin{equation}
\Phi = \alpha(t) -  \exp[-\frac{k}{d}t]\int \dot\alpha(t')\exp[-\frac{k}{d}t']dt' . 
\end{equation}
In the limit $ t \rightarrow \infty $, we have a simple expression $ \Phi(t) \simeq \alpha(t) $, 
from which one sees that the center of the DW is given by 
\begin{equation}
R(t) = \frac{A}{B}\alpha(t) + R_0 
\label{Ralpha}
\end{equation}
with the initial position $ R_0 $.  This result shows a quite simple meaning; the motion of the DW can be directly 
controlled by the modulation of the  external magnetic field. That is, the angular modulation just gives rise to the 
motion of the center of the DW. 

This result  (\ref{Ralpha})  simply reveals  an interplay between the geometric action, which is characterized by the 
coefficient $ A$, and the dissipative function that is described by $ B $. in the asymptotic limit: 
$ t \rightarrow \infty $.  In the intermediate stage,  the other term characterized by the coefficient $ D $plays a role 
to disappear asymptotically. 

As a special case that there is  no dissipation, namely, $ B $ and $ D $ vanish, we see that $ \dot \Phi = 0, ~~\dot R = k \sin( \Phi- 
\alpha(t) ) = 0 $.  Here we examine the simple case: $ \alpha = \omega t $, for which one sees that a sinusoidal oscillation behavior 
$ R(t) = \frac{1}{\omega} \sin(\Phi_0 - \omega t) $.  This is contrast to the simple linear behavior, which is just caused by the 
dissipative effect. 

In this way the mechanism of the steering of the DW obeys the same mechanism as the  {\it self -propulsion of swimmer}  or 
the so-called "falling cat", which explains the creation of rotation (angular momentum) as a result of the deformation of moving 
body\cite{geometric}. This kind of phenomena is an example of classical ``holonomy". 
.

{\bf Case (II)} : The amplitude $ b $ is allowed to be time varying. 
For this purpose it suffices that one restricts the case that the phase $ \alpha $ is constant, actually zero, hence 
the equation of motion for $ \Phi $ turns out to be 
\begin{equation}
(D+\frac{A^2}{B})\dot\Phi = -k(t) \sin \Phi, 
\end{equation}
which is solved to be 
\begin{equation}
\sin\Phi(t) = \frac{1}{\cosh \tilde K(t)}, 
\end{equation}
where $ \tilde K(t) = \int^t \tilde k(t)dt $ with $ \tilde k(t) = k(t)/d $. 
By changing the time variable by the relation $ d\tau = \tilde k(t)dt $, we see that the DW center moves 
according to the relation: 
\begin{align}
 R(\tau) & = \frac{A}{B} \int^{\tau}_{\tau_0} \frac{d(\sinh\tau)}{\cosh^2\tau} \nonumber \\
              &  = -\frac{A}{B}\big[ \tan^{-1}(\sinh\tau_0)  - \tan^{-1}(\sinh\tau)\big] , 
\end{align}
which tends to the asymptotic value $ \frac{A}{B}\big[\frac{\pi}{2} -  \tan^{-1}(\sinh\tau_0)]   $ 
for $ t \rightarrow  \infty $.  This feature is essentially different from the control by the modulation of 
the phase $ \alpha(t) $ (\ref{Ralpha}).  \\

{\it Physical observability}: 
The steering of the kink may be observable by using the induction effect caused by the 
change of magnetic flux.  Namely let us consider a conducting loop which is arranged such that 
the one-dimensional spinor BEC {\it perpendicularly}  penetrates through the center of the loop. By this geometry, 
the magnetic flux, denoted by $  \Sigma $, can be calculated as $ \Sigma \propto S_x {\cal A} $ ( $ {\cal A}  $  means the area of the loop).  
This  may be averaged over some interval on the x-axis to give the form $ \Sigma(t)  \propto \cos(\eta(0)\Phi(t)) $,according to 
the procedure given above.  Hence it is possible to  derive  the induced (electric) current due to the Faraday effect, that is 
$ \dot\Sigma \propto  \dot\Phi $.  By observing this current, one may have an evidence of the steering of the domain wall 
via the phase $ \Phi $,  together with $ R $. 
\\

{\bf Remark on the kinetic energy}: 
As is implied from the Lagrangian in (\ref{LC}), $ (R, \Phi) $ forms a pair of canonical variables. 
About this, we give a brief comment. Our concern is the connection with the fluid kinetic energy $ K_{\phi} $. This can be written as 
\begin{equation}
  K_{\phi} = \int \sin^2\frac{\theta}{2} (\frac{d\eta}{dx})^2 \Phi^2 dx 
  = \frac{\hbar^2}{2m}F_0^2\Phi^2, 
  \label{Phi2} 
  \end{equation}
which is of the second order $ \epsilon $ compared with the (\ref{canon}) in the adiabatic approximation. 
This term can be clarified by considering the magnetic field written as $ {\bf H} = (\frac{h(t)}{2} x^2, 0, 0 ) $, namely, the parabolic form 
confinement potential with time-modulating coefficient written in the form: 
\begin{equation}
\int \Psi^{\dagger} ({\bf \sigma}\cdot {\bf H}\Psi )dx = \frac{h(t)}{2}\int \sin\theta \cdot x^2 \cos[\eta(x-R(t)\Phi(t)] dx, \nonumber
\end{equation}
 which yields, by using the  localization property of $ \sin\theta $.  
\begin{equation}
V_{\phi} \simeq \frac{h(t)}{2} \cos [\eta(0)\Phi] R^2. 
\end{equation}
Thus we have the effective Hamiltonian 
\begin{equation}
{\cal H}_{eff} = \frac{M}{2}\Phi^2 + \frac{h(t)}{2} \cos \Phi R^2, 
\end{equation}
which surely suggests that the pair $ (R,\Phi) $ gives a canonical pair; namely, $ R  $ represents the 
"coordinate "  and " $ \Phi $ its conjugate momentum. This feature is very different from the 
case for the first topic, for which we have no angle variable $ \Phi $.  Specifically the time 
dependence of the coefficient $ h(t) $: If this is arranged  such that $ h(t) \simeq \sum \delta(t- nT) $ 
( $ n = $ integers),  we have a problem that is analogous to the problem {\it kicked oscillator} \cite{Zaslavsky}.  
The details may be left for future study. \\

\section{Summary}

We have studied two problems for the dynamics of the domain wall. 
The first one concerns  the collective dynamics of the center of the DW, which 
treats purely the translational motion for the kink resulting in the one-dimensional 
quantum potential model.  The second one is rather intricate compared with the first one; we discussed the coupled dynamics for 
the translational motion of the DW and the angular mode inherent in the spin degree of freedom in 
the BEC: As a consequence of the interplay between the canonical term and the dissipative effect, the 
control (steering) for the DW motion can be achieved. 

Our starting point is the two-component wave for the spinor BEC, which may be regarded as the simplest case of the the spin coherent state 
\cite{HK1,Arecchi}. From this point of view, the present formalism is straightforwardly extended to the spinor condensates with higher spin. 
As other topic for future study, mentioned is the fluctuation besides the dissipation \cite{Hanggi}, which is a complement 
as inferred from a fluctuation dissipation theorem and  inevitable in dealing with the spinor condensates that incorporate the random impurities .   
The similar problem dealing with the random effects has been given  in the vortex motion in the second kind superconductivity (see, e.g.,  
\cite{Dorsey}).  \\

The author would like to thank Dr.S.Tsuchida for preparing the figure.

\end{document}